\documentclass[a4paper,aps,showpacs,showkeys,prm,reprint,superscriptaddress]{revtex4-1}
\usepackage{amsmath,multirow,dcolumn,float}
\usepackage[american]{babel}
\usepackage[T1]{fontenc}
\usepackage{natbib}
\usepackage[utf8]{inputenc}
\usepackage{graphicx}
\usepackage{epsfig}
\usepackage{xcolor}
\usepackage{subfigure}
\usepackage{dcolumn}
\usepackage[normalem]{ulem}
\usepackage{csquotes}
\usepackage{color}

\definecolor{gruen}{rgb}{0,0.4,0}
\newcommand{\ch}[1]{#1}

\begin{document}

\title{
  \ch{Fermi level pinning by defects can explain the large reported
  carbon 1s binding energy variations in diamond.}
}
\date{\today}

\author{Michael \surname{Walter}}
\email{Michael.Walter@fmf.uni-freiburg.de}
\affiliation{Fraunhofer IWM, MikroTribologie Centrum $\mu$TC, Wöhlerstrasse 11, D-79108 Freiburg, Germany}
\affiliation{FIT Freiburg Centre for Interactive Materials and Bioinspired Technologies, University of Freiburg, Georges-K\"ohler-Allee 105, 79110 Freiburg, Germany}
\affiliation{Institute of Physics, University of Freiburg, Herrmann-Herder-Straße 3, D-79104 Freiburg, Germany}
\author{Filippo \surname{Mangolini}}
\affiliation{
Materials Science and Engineering Program and Department of Mechanical Engineering, The University of Texas at Austin, Austin, Texas 78712, USA
}
\author{J. Brandon \surname{McClimon}}
\affiliation{Department of Materials Science and Engineering, University of Pennsylvania, Philadelphia 19104, USA
}
\author{Robert W.  \surname{Carpick}}
\affiliation{Department of Mechanical Engineering and Applied Mechanics, University of Pennsylvania,
Philadelphia, Pennsylvania 19104, USA}
\author{Michael \surname{Moseler}}
\affiliation{Fraunhofer IWM, MikroTribologie Centrum $\mu$TC, Wöhlerstrasse 11, D-79108 Freiburg, Germany}
\affiliation{Freiburger Materialforschungszentrum, Universität Freiburg, Stefan-Meier-Straße 21, D-79104 Freiburg, Germany}
\affiliation{Institute of Physics, University of Freiburg, Herrmann-Herder-Straße 3, D-79104 Freiburg, Germany}
	
\begin{abstract}
  The quantitative evaluation of the carbon hybridization state by
  X-ray photoelectron spectroscopy (XPS) has been a surface-analysis
  problem for the last three decades due to the challenges associated
  with the unambiguous identification of the characteristic binding
  energy values for sp$^2$ and sp$^3$-bonded carbon. 
  While the sp$^2$ binding energy is well established, there is 
  disagreement for the sp$^3$ value in the literature.
  Here, we compute the binding energy values for model structures
  of pure and doped-diamond
  using density functional theory.
  The simulation results indicate that the large band-gap of diamond
  allows defects to pin the Fermi level, which results in large
  variations of the C(1s) core electron energies for
  sp$^3$-bonded carbon, in agreement with the broad range of
  experimental C(1s) binding energy values for sp$^3$ carbon
  reported in the literature.
  Fermi level pinning by boron is demonstrated 
  by experimental C(1s) binding energies of highly B-doped
  ultrananocrystalline diamond that are in good
  agreement to simulations.
\end{abstract}


\maketitle

\ch{\section{Introduction}}

X-ray photoelectron spectroscopy is one of the most 
powerful tools for the characterization of carbon-based materials 
\cite{Chu06mcp}. 
The analysis of the bonding configuration of carbon is normally 
carried out by XPS through the acquisition of carbon 1s (C(1s)) spectra. 
The spectra are often fitted with two distinct features, one assigned 
to threefold-coordinated (sp$^2$) carbon and one assigned to 
fourfold-coordinated (sp$^3$) carbon. 
In spite of the wide use of this analytical procedure for the 
quantitative evaluation of the carbon hybridization state 
on the basis of C(1s) XPS signals, the validity of this methodology 
has recently been questioned \cite{Mezzi10sia,Kaciulis12sia} and 
still remains a matter of discussion in the literature \cite{Fujimoto16ac}.

We have recently shown that absolute XPS peak positions
can be predicted by facile spin-paired DFT calculations
within the frozen core approximation \cite{Walter16prb,Walter19pccp}.
In this approach, an empirical shift appears that
can be obtained from experimental gas-phase XPS data.
The main difficulty in assigning core hole energies 
in general, and C(1s) energies in particular, is the
unambiguous definition of the reference energy.
This is not the case for gas-phase investigations, where
the energy of an electron in vacuum far away from
the mother ion can be used as a
well-defined reference for
the core electron binding energy $E_B$.

For solid samples the Fermi energy is usually used
as the reference energy level,\cite{Riga77mp} but this
energy can only be defined correctly for systems without
a band gap, i.e., metals or semi-metallic systems like
graphite.
Accordingly, the experimental variation in C(1s) energies
of semi-metal graphite is moderate and found to range from
284.28 eV to 284.63 eV \cite{J._Keiser1976,Morar86prb,Lascovich91ass,Waite92apl,Y._Xie1992,G._Witek1996,Z._Bastl1995,R._Graupner1998,T.Y._Leung1999,Bobrov01PRB,Kusunoki01ss,Yan04jpd,Saw04ml,Ghodbane10drm,Mezzi10sia,Kaciulis12sia,Schenk16jpc}.
Diamond, in contrast, has a very large band gap (its experimental
value is 5.5 eV \cite{Ashcroft76}), which hinders the
consistent definition of an energy reference. 
Additional difficulties arise from the presence of defects, 
impurities, and dopants in diamond 
(such as nitrogen or boron\cite{Kaiser59pr,Dannefaer07pss,Himpsel79prb}), 
which may pin the Fermi energy.
These factors, together with effects of surface charging when 
acquiring XPS spectra on diamond, have resulted in the publication 
of a wide range of binding energy values for the 
C(1s) signal of sp$^3$-bonded carbon (in the range
of 283.25 eV - 291.35 eV)
\cite{R._Graupner1998,T.Y._Leung1999,K.G._Sawa2004,Kawabata_Yusaku2004}.
Despite the difficulties in unambiguously assigning a characteristic 
binding energy value to sp$^3$-bonded carbon, the quantitative evaluation 
of the carbon hybridization state on the basis of C(1s) XPS signals 
is still widely performed on the basis of fitting the C(1s) XPS spectra 
with two synthetic peaks, one assigned to sp$^2$-hybridized carbon 
and one to sp$^3$-hybridized carbon.
Here, we use DFT calculations to compute the characteristic 
C(1s) binding energy values for sp$^3$-bonded carbon in diamond and 
sp$^2$-bonded carbon in graphite. 

\ch{\section{Methods}}

Our DFT calculations were carried out 
with the GPAW
\cite{Mortensen05prb,Enkovaara10jpc} package,  an implementation of
the projector augmented wave (PAW) 
method \cite{Bloechl:PAW}.
The PAW method splits the Kohn-Sham
wave functions into a smooth part, representable on
configuration or momentum space grids, and corrections that are local
near to the atoms. We use the
configuration space grid implementation, and apply a grid
spacing of 0.2 {\AA} for representing the smooth part of the
Kohn-Sham wavefunctions,
unless noted otherwise. 
The exchange correlation energy was approximated by the
generalized gradient correction as proposed
by Perdew, Burke and Ernzerhof (PBE)\cite{Perdew96prl}. 
\ch{We have used a spin-paired description for simplicity 
  of band-structure analysis presented below. 
  We have checked that the inclusion
  of spin does not change our results significantly.}  
The structures were relaxed until all forces on the atoms dropped 
below 0.05 eV\text{\AA}$^{-1}$.

Lower energy atomic states can be conveniently 
held fixed in their atomic form within the frozen core
approximation, where we include the $1s$ electrons for C
and the diamond dopant atoms B, N, and the
$1s,2s, 2p$ electrons for the diamond
dopant atom P in our calculations.
A similar approach is adopted to describe the core hole
by lowering the occupation of the relevant state in the
atomic calculation by one. The resulting \ch{self-consistently obtained}
Kohn-Sham orbitals \ch{in the presence of the core hole}
are then used to construct the frozen core
\cite{Enkovaara10jpc,Ljungberg11jesrp,Susi14bjn}.

\begin{figure}
  \centering
  \includegraphics[width=0.49\textwidth]{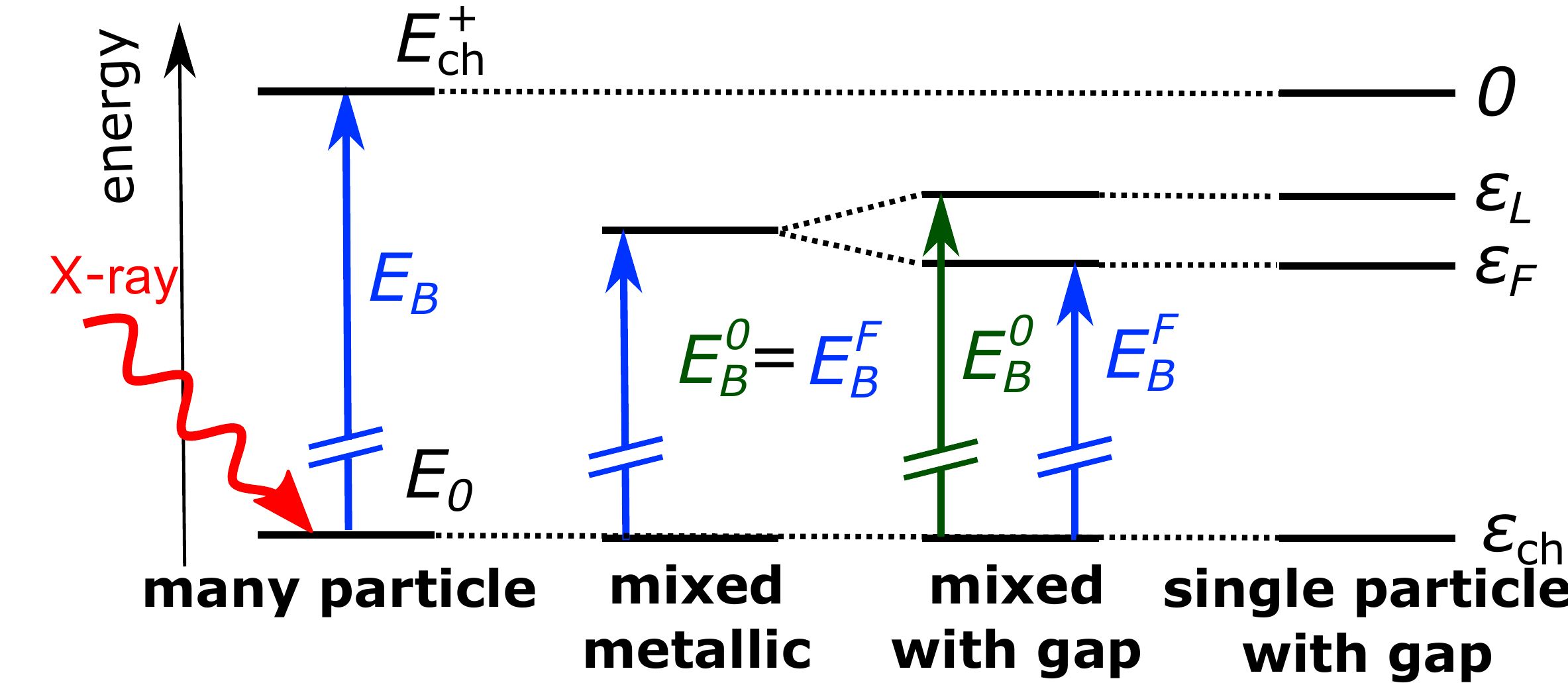}
  \caption{
    \ch{Schematic of many particle (capital letters) and single particle 
      (small letters) energies involved in
      the definition of the core-hole binding energy 
      corresponding to Eqs.~(\ref{eq:ech}-\ref{eq:EB0}).}
 }
 \label{fig:energies}
\end{figure}
\ch{The calculation of core-hole binding energies
  follows the methodology developped in Ref.~\onlinecite{Walter16prb},
  as illustrated in Fig.~\ref{fig:energies} and is
  briefly summarized in what follows.}
XPS energies of molecules or clusters
in the gas phase are well defined by the many body
ground state energy $E_0$ and
the core hole excited energy $E_{\rm ch}^+$.
The latter can be approximated as the ground
state of the system with a core hole in the frozen core, where
the self-consistent calculation of the valence
orbitals in the field of the core hole automatically
takes into account relaxation effects.
The photoelectrons' binding energy is then:
\begin{equation}
  E_B=E_{\rm ch}^+-E_0 \; .
  \label{eq:ech}
\end{equation}
XPS energies for 
periodic systems as modeled here have to be treated in
a mixed many-body and single-particle picture.
Experimentally, the core level
energy is measured relative to the single particle Fermi level
$\varepsilon_F$ \ch{(smaller than zero)} that is aligned
to the Fermi level of the detector by holding them at a common
ground\cite{Riga77mp,Ozaki17prl}:
\begin{equation}
  E_B^F= E_{\rm ch}^+ - E_0 + \varepsilon_F \;.
  \label{eq:EBF}
\end{equation}
The computational difficulty of charged super-cells \cite{Bruneval14prb}
for the description of the core ionized final state
can be overcome by neutralizing the super-cell by an extra electron.
This electron will locate itself
in the lowest unoccupied ``molecular'' orbital (LUMO).
Neglecting all other contributions due to the change in
the electron density, this approach
adds the energy $\varepsilon_L$ to $E_{\rm ch}^+$. 
We obtain a binding energy in the neutralized system of
the form:
\begin{equation}
  E_B^0=(E_{\rm ch}^+ + \varepsilon_L)-E_0 = 
  E_B^F + \varepsilon_L - \varepsilon_F \; .
  \label{eq:EB0}
\end{equation}
For a system without a band gap, $\varepsilon_F=\varepsilon_L$ and
hence conveniently $E_B^F=E_B^0$.
Systems with a band gap pose several problems, however.
The position of the Fermi level $\varepsilon_F$ is not defined 
in these systems which prohibits a defined value for $E_B^F$.

\ch{\section{Results}}

In the first step, we consider the case of bulk 
diamond, which is the most problematic system due to its large band-gap.
We model diamond by unit cells containing 64 atoms
in tetrahedral configurations
with the experimental lattice constant of
3.5669 \AA\cite{CRC04} resulting in a CC bond-length of 1.55 \AA.
We are particularily interested in the effect of
defects and impurities on the expected C(1s) value.

One of the most common impurities in natural diamond gemstones is
nitrogen\cite{Kaiser59pr,Dannefaer07pss}
which can be present in amounts of up
to 0.5 at.~\% \cite{Lombardi03jpc}.
Different possibilities for the presence of nitrogen are considered
in the literature, where the replacement of a C atom with a N
atom with or without
a neighboring vacancy are believed to be the most common
configurations\cite{Goss04drm}.
In the former case, the P1 center\cite{Iakoubovskii00jpc,Lombardi03jpc}
is modeled as the replacement of a single C atom by N (C$_{63}$N).
The nitrogen vacancy (NV) center\cite{Jelezko06pss} is modeled 
by the replacement of a single C atom by N and a
neighboring vacancy (C$_{62}$N).
The A center \cite{Goss04drm} is two neighboring N atoms
(C$_{62}$N$_2$),
and the N3 center\cite{Goss04drm,Lu12ac} consists of
three nitrogen atoms near to a vacancy (C$_{60}$N$_3$). 
The B center\cite{Goss04drm} is
four N surrounding a vacancy (C$_{59}$N$_4$).

Other common impurities are boron or phosphorus\cite{Koizumi00drm}.
These elements are either present in natural
diamonds (type IIb in the case of boron\cite{Morar86prb,Bobrov01PRB})
or introduced artificially\cite{Bobrov01PRB,Kusunoki01ss,Ghodbane10drm}.
Boron- or phosphorus-doped structures are modeled by replacing
a single C atom by B (C$_{63}$B) or P (C$_{63}$P), respectively.
All model structures were relaxed to their next local minimum
while the unit cell was kept fixed.
\ch{Relaxed structures and distributions of calculated C(1s) values
  are depicted in Supplemental Material.}

\begin{figure}
  \includegraphics[width=0.5\textwidth]{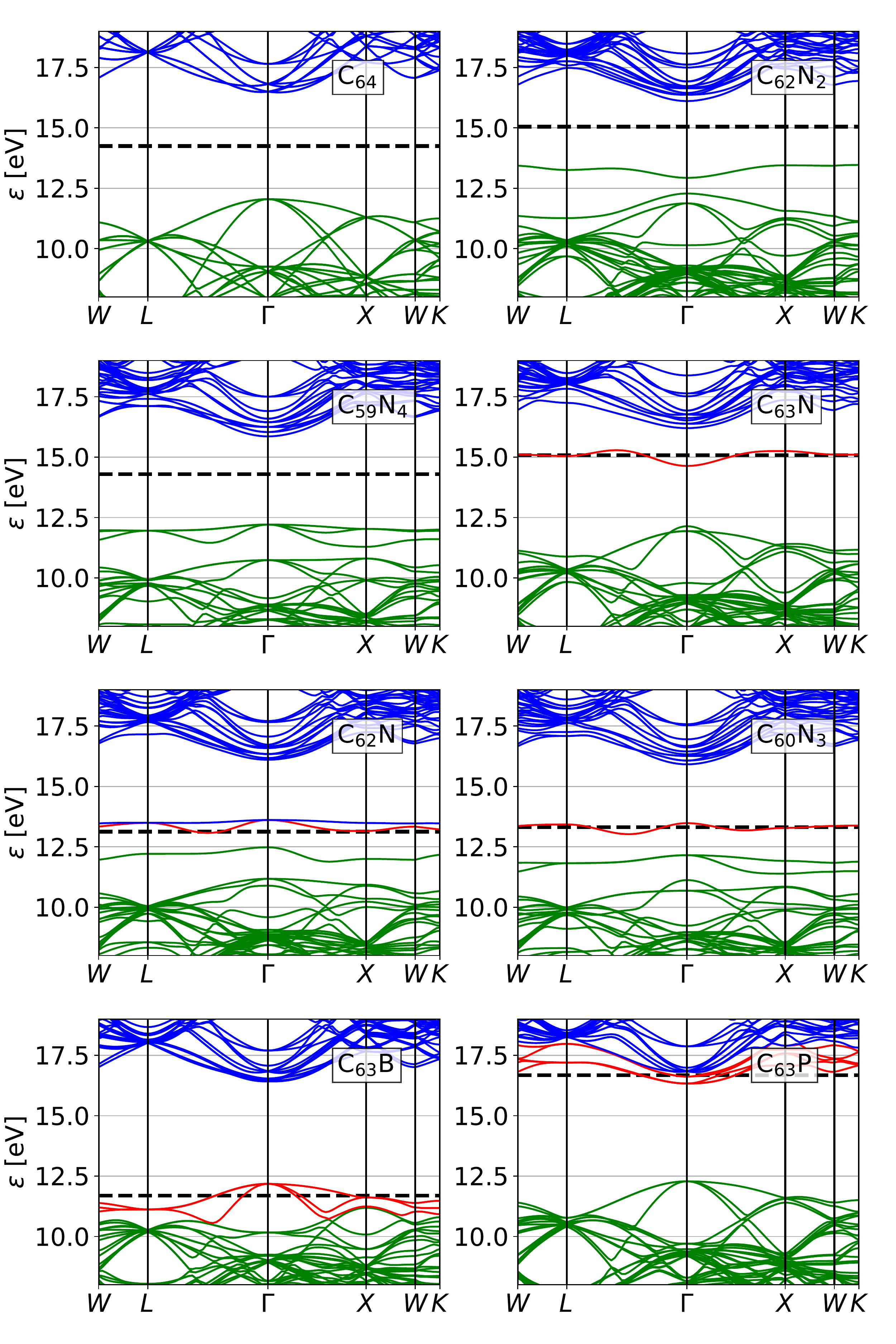}
  \caption{
    Band structure of diamond, pure and with defects.
    Fully occupied bands are colored in green,
    partly occupied bands in red and empty bands in blue.
    The broken line shows the Fermi level $\varepsilon_F$.
 }
 \label{fig:bandC64}
\end{figure}
Fig.~\ref{fig:bandC64} shows the effect of defects
on the diamond band structure.
While the overall band-structure is rather similar in all cases,
the defects or dopants introduce states at different
positions within the diamond band gap. 
The gap is found to be 4.42 eV wide
according to PBE\cite{Matsuda10jpcl,Yang13jcp,Rivero16carbon}
underestimating the experimental value of
$\sim$5.5 eV.\cite{Ristein06ss,Oyama09apl}
This well known tendency of local and semilocal functionals
to understimate the gap can be improved by computationally
more demanding hybrid functionals\cite{Yang13jcp,Rivero16carbon},
but will not affect the general effect of Fermi level pinning
discussed in the following.

The two N in C$_{62}$N$_2$ slightly disturb a high 
lying unoccupied (acceptor) levels near to
the conduction band edge $\varepsilon_c$ and introduce 
a rather high occupied (donor) level above the valence band edge
$\varepsilon_v$\cite{Goss04drm}, thus reducing the band gap.
These levels show some variation with the wave vector in our calculation
indicating interaction between periodic images.
\ch{The variations are rather small (a few 100 meV) and are neglected
  in the following.}
A similar effect as in C$_{62}$N$_2$ is observed in C$_{59}$N$_4$, where mainly
states slightly above $\varepsilon_v$ are introduced.
Similar to pure diamond,  the Fermi level
$\varepsilon_F$ is located in the middle between the maximum of
the highest occupied and the minimum of the lowest unoccupied band.
An infinitesimal charge could move the Fermi energy towards
the valence or the conduction band depending on its sign.

The freedom of moving the Fermi level by slight charging
is not the case anymore for the other models presented in 
Fig.~\ref{fig:bandC64} as these contain an
unpaired electron.
There, $\varepsilon_F$ is pinned by this half-filled state
and is therefore well defined.
The exact positions of the
Fermi level relative to the band structure of diamond
varies largely depending on the nature of the defect, however.
While the half filled state is close to the conduction band in
C$_{63}$N, it is found in the middle of the band gap for
C$_{62}$N and C$_{60}$N$_3$. The most extreme positions are found
in C$_{63}$B where the partially filled states 
(degenerate at the $\Gamma$-point) are
near to $\varepsilon_v$\cite{Lee06prb,Oyama09apl} and for C$_{63}$P 
where partially filled states are found near to $\varepsilon_c$.\cite{Oyama09apl}

Assuming that the C(1s) core level energy relative to the vacuum level
$E_B$ is mainly constant for carbon atoms far from
any defect, the variation in $\varepsilon_F$ will be reflected
in a variation in $E_B^F$ [c.f.~eq.~(\ref{eq:EBF})],
which is indeed the case as will be shown now.

\begin{figure}
  \centering
  \includegraphics[width=0.5\textwidth]{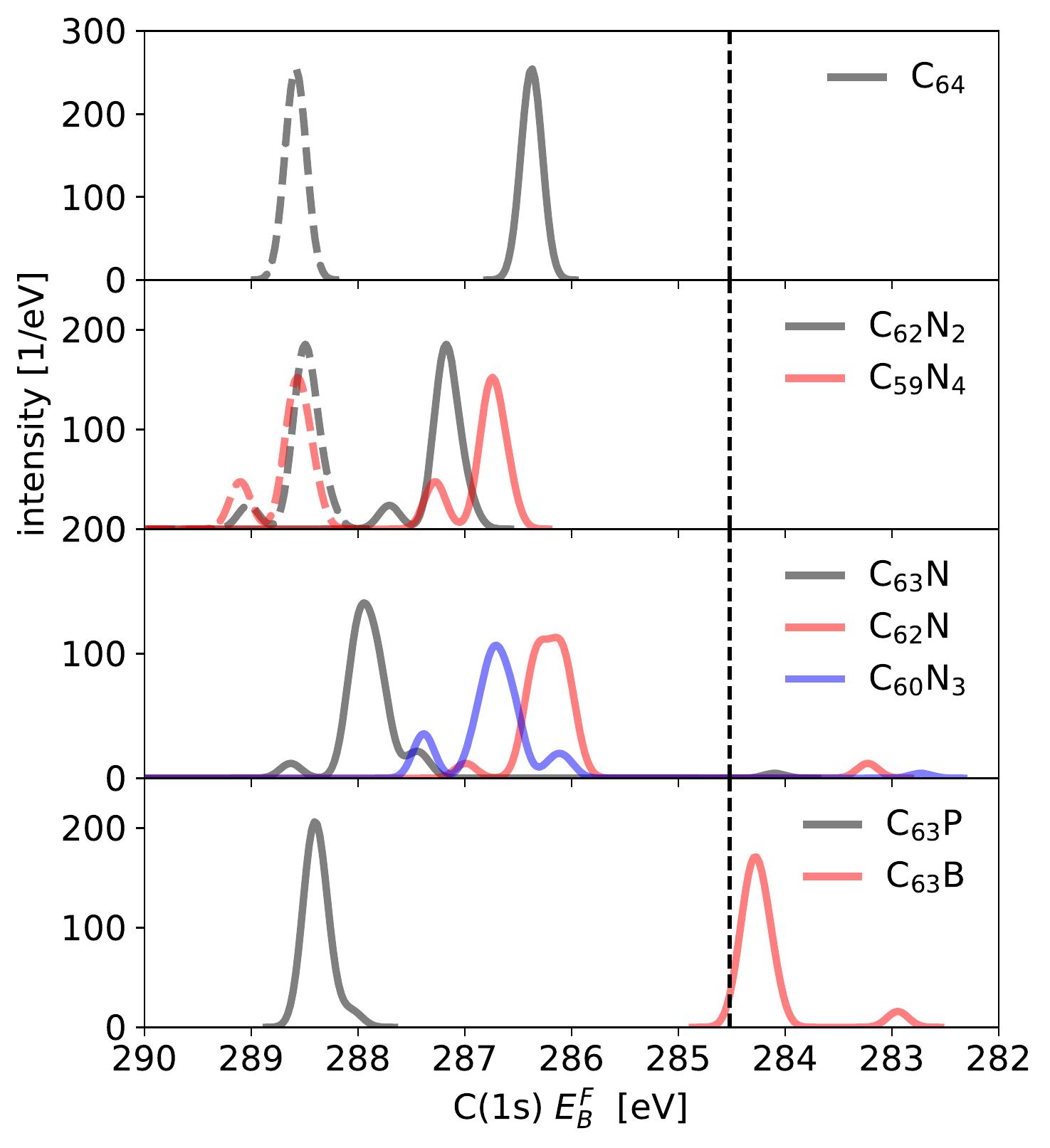}
  \caption{
    Absolute XPS spectra for the models derived from the C$_{64}$
    models without and with defects and impurities.
    The broken line is without the correction to $\varepsilon_F$
    (i.e., $E_B^0$) and
    the solid line includes this shift (i.e., $E_B^F$).
    The spectra are obtained
    by convoluting with Gaussians of 0.24 eV full width at half maximum (FWHM).
    The broken vertical line indicates graphite C(1s) energy 
    \ch{\cite{Walter16prb}}.
  }
  \label{fig:XPS64}
\end{figure}
The C(1s) XPS spectra corresponding to the defects are compared to
pure diamond in
Fig.~\ref{fig:XPS64}. The unshifted spectra
[broken lines, corresponding to $E_B^0$ in eq.~(\ref{eq:EBF})]
of the C$_{64}$, C$_{62}$N$_2$ and C$_{59}$N$_4$  models are very similar
in their main peak as they share the same lowest
unoccupied level (c.f.~Fig.~\ref{fig:bandC64}).
The spectra are different when they are corrected to $\varepsilon_F$
located at the center of the gap
\ch{(as usually assumed in semiconductors \cite{Yu10})}
due to the different 
band gap values.
Unshifted $E^0_B$ and shifted $E_B^F$
coincide for the cases involving an unpaired electron
as $\varepsilon_L=\varepsilon_F$ in these cases.
The largest core level energy is found for C$_{63}$N and
the lowest energy for C$_{63}$B
as expected from the positions of the unpaired levels in
Fig.~\ref{fig:bandC64}.
The main C(1s) energy in C$_{63}$B of 284.27 eV is in good agreement
to the value of 284.4 eV measured for artificially
boron-doped diamond\cite{Kusunoki01ss} and
to the main peak of strongly boron-doped ultranano-crystalline 
diamond (UNCD) found at 284.09$\pm$0.05 eV discussed in detail below.

\begin{table}
  \begin{tabular}{lllll}
    structure & model        & gap [eV] & C(1s) [eV] & shift [eV]\\
    \hline
    pure      & C$_{64}$      & 4.42 & 286.36\ch{$^*$} &  1.85 \\
    P1 center & C$_{63}$N     & 0    & 288.41 &  3.43 \\
    NV center & C$_{62}$N     & 0    & 286.15 &  1.68 \\
    A center  & C$_{62}$N$_2$ & 2.64 & 287.17\ch{$^*$} &  2.65 \\
    N3 center  & C$_{60}$N$_3$ & 0   & 286.70 &  2.18 \\
    B center  & C$_{59}$N$_4$ & 3.65 & 286.74\ch{$^*$} &  2.22 \\
    B         & C$_{63}$B     & 0    & 284.27 & -0.25 \\
    P         & C$_{63}$P     & 0    & 288.41 &  3.89 \\
    graphite  & C$_{72}$      & 0    & 284.52\cite{Walter16prb}
  \end{tabular}
  \caption{Pure and defected diamond model structures, their PBE band gaps,
    the position of their main C(1s) peak \ch{($E_B^F$)} 
    and the resulting shift
    relative to graphite C(1s) (see also Fig.~\ref{fig:XPS64}).
    \ch{The C(1s) values marked by an asterisk are corrected assuming
      the Fermi level to lie in the center of the band gap.}
  }
\end{table}
Various side peaks appear for the different defects.
In C$_{62}$N$_2$, there is a peak at slightly higher energy 
that comes from the six carbon atoms surrounding the two nitrogen atoms
forming a N(CR$_3$)$_3$ like structure with an 
NC bond length of 1.45 {\AA} very similar to the 1.46 {\AA}
in N(CH$_3$)$_3$.
In contrast, a lower energy peak from the four carbon atoms
surrounding N is found in C$_{63}$N. Here, the N has to share
the bonds with four C due to symmetry.
The NC bond-length is thus much longer (1.60 {\AA})
exceeding the diamond CC bond.

For the N impurity with a neighboring vacancy (C$_{62}$N), the
lower energy peak (283.2 eV) 
is from the three carbons surrounding the cavity and 
the higher energy peak (287 eV) 
from the three carbons connected to nitrogen.
The NC bond-length of 1.48 {\AA} is quite similar to N(CH$_3$)$_3$ again.

The four carbon atoms surrounding the P atom symmetrically
with a bond length of 1.70 {\AA} in C$_{63}$P  produce the
lower energy shoulder in the corresponding C(1s) spectrum.
Similarly, the lower energy peak around 283 eV in C$_{63}$B
is from the four carbon atoms surrounding B.
Structurally, the B impurity atom shares four equal bonds with large
CB bond length of 1.59 {\AA} as compared to 1.55 in CC.

Besides these side peaks the main effect is the strong variation of the
C(1s) peak with defect type. A defect in diamond
dictates the C(1s) position of E$_B^F$ even for carbon atoms far apart
due to its influence on the Fermi level.
Therefore we can conclude that diamond is a problematic
system to define the sp$^3$ C(1s) energy as the
C(1s) energy depends on the 
nature of defects that are omnipresent in real diamonds.
This finding can contribute explaining the broad range of C(1s) values for
diamond reported in the 
literature.\cite{Fujimoto16ac}

We finally seek for an experimental validation of the computional
analysis presented so far. 
Experimental XPS data were acquired on hydrogen-terminated 
ultrananocrystalline diamond (UNCD Aqua 25, Advanced Diamond Technologies, 
Romeoville, IL, USA), boron-doped ultrananocrystalline diamond 
(UNCD Aqua 25, Advanced Diamond Technologies, 
Romeoville, IL, USA),\cite{Zeng15carbon} and
freshly cleaved highly ordered pyrolitic graphite 
(HOPG, grade 2, SPI Supplies, West Chester, PA, USA).
Near-edge X-ray absorption fine structure (NEXAFS) spectroscopy
measurements indicated that the fraction of
sp$^3$-bonded carbon in undoped and boron-doped UNCD was 94+/-3\% and
96+/-3\%, respectively \cite{Mangolini14ac,Zeng15carbon,Mangolini16ac}.
In the present work, the X-ray source was run at 30 mA and 12 kV,
whereas the analyzer was operated in constant-analyzer-energy (CAE) mode.
Survey spectra were acquired with the pass energy and step size equal
to 200 eV and 1 eV, respectively. For the high-resolution (HR) spectra,
the pass energy and step size were, respectively, 100 and 0.05 eV
(full width at half maximum (FWHM) of the peak height for the
Ag 3d$_{5/2}$ equal to 0.57 eV).
The curved slit at the entrance of the hemispherical analyzer
has a width of 0.8 mm.
The residual pressure in the analysis chamber was always below
10$^{-6}$ Pa.  The spectrometer was calibrated according to
ISO 15472:2001 with an accuracy better than $\pm$0.05 eV.
All the XPS results reported here are mean
values calculated from at least three independent measurements,
with the corresponding standard deviation reported.

\begin{figure}
  \includegraphics[width=\linewidth]{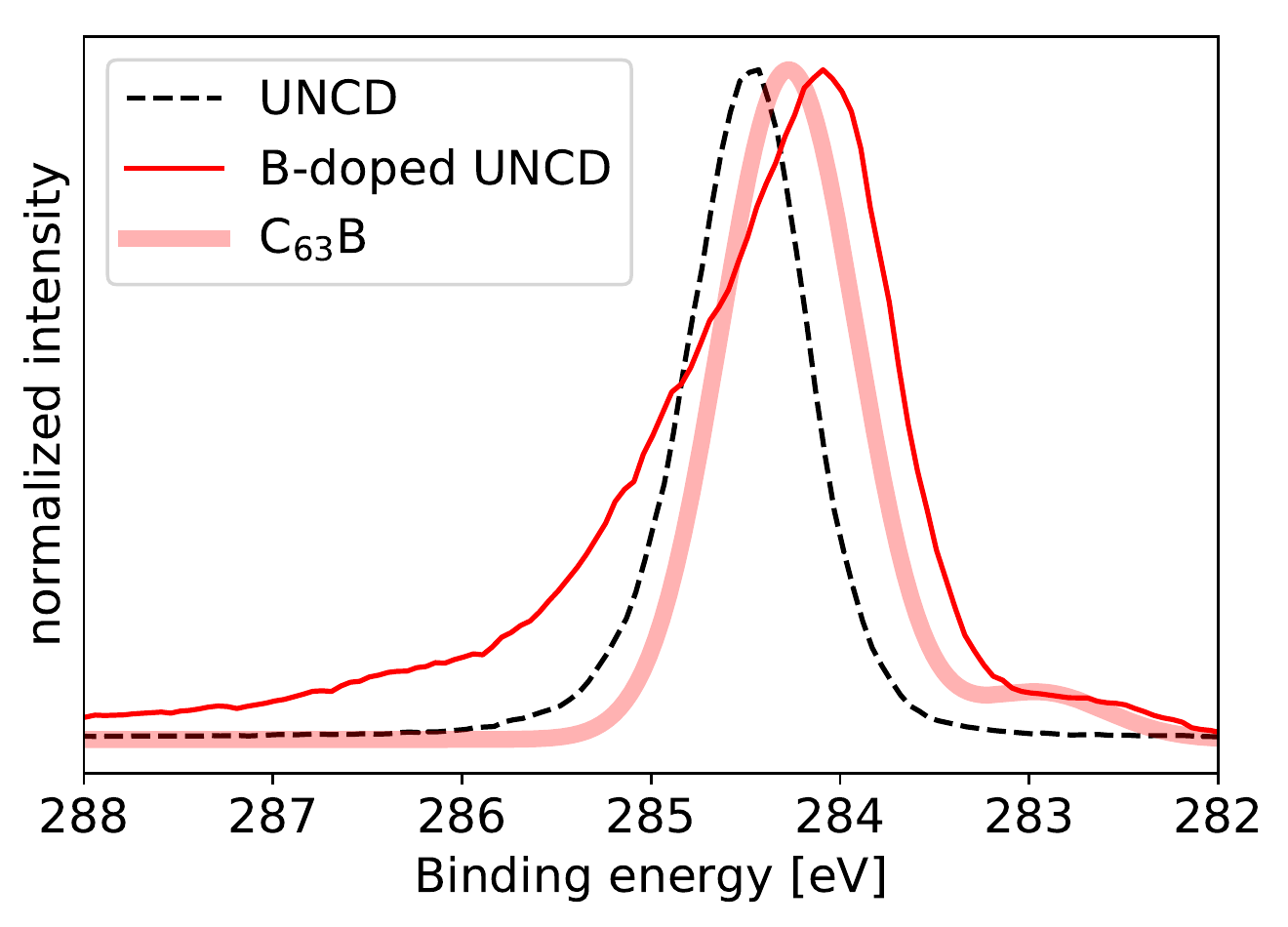}
  \caption{
    \label{fig:BoronUNCD}
    Experimental C(1s) spectrum of ultranano-crystalline diamond 
    (UNCD) and B-doped UNCD 
    compared to the calculated spectrum of the C$_{63}$B model.
    The calculated C(1s) energies are convoluted by Gaussians of 0.82 eV FWHM.
  }
\end{figure}
The DFT calculations are in good agreement with experimental 
XPS data acquired on undoped and B-doped UNCD  
shown in Fig.~\ref{fig:BoronUNCD}.
First, the binding energy of the characteristic C(1s) peak for 
B-doped UNCD is lower than the binding energy of the C(1s) 
signal of undoped UNCD (284.09$\pm$0.05 eV vs. 284.47$\pm$0.05 eV, 
respectively in experiments) and agrees well with
the 284.27 eV from DFT.
Second, as in the simulation, there is a clear shoulder on the 
lower binding energy side in the experiment. 
Note that the shoulder on the high binding energy side of the 
experimental spectrum (286-288 eV) is caused by C-O 
bonds\ch{\cite{Arezzo94,Dementjev97,Kaciulis12sia,Mezzi10,Retzko03,Wilson01,Yang11}} 
in the 
near-surface region.
\ch{The experimental C(1s) of undoped UNCD 
  does not agree with the 
  calculation for pure diamond. This is not surprising with respect to the 
  large band gap of diamond and the resulting undefined Fermi level.
  UNCD is not single-crystal diamond and the grain boundaries contain
  sp$^2$-carbon, defects, and are rich in hydrogen.
  The effect of hydrogen on the shape and position of C(1s)
  signal of carbon materials will be presented
  in a later publication.
  The strong influence of these defects is also inline
  with the observation that etching can shift the C(1s) 
  value by several eV in UNCD films \cite{Sumant07prb}.
}

\ch{\section{Conclusions}}

In conclusion, we have shown that the sp$^3$ C(1s) binding energies determined
from diamond are highly affected by the presence and nature of defects.
This strong dependence of the binding energy of the characteristic
C(1s) peak for sp$^3$ carbon on the type and number density of defects
in diamond samples makes the use of diamond as reference material
for XPS analysis potentially misleading.
This is a consequence of the large gap, i.e. the insulating 
nature,\cite{Kaciulis12sia} of
undoped carbon and the resulting absence of a 
defined refence energy within the system.
\ch{This conclusion is not restricted to diamond, but applies for
  XPS measurements of other wide band gap materials, where similar 
  spreads in experimental core hole binding energies have been 
  observed as discussed in Ref.~\onlinecite{Walter16prb}.} 

\begin{acknowledgments}

M.W. and M.M. thank G. Moras for useful discussions.
Computational resources of FZ-Jülich and NEMO are thankfully acknowledged. 
R.W.C. acknowledges support from the U.S. National Science Foundation
(NSF) through the University of Pennsylvania Materials Research
Science, and Engineering Center (MRSEC) (DMR-1720530).
F.M. acknowledges support from from the Welch Foundation under
Grant F-2002-20190330, the Marie Curie International 
Outgoing Fellowship for Career Development within the 7th 
European Community Framework Programme under contract no. 
PIOF-GA-2012-328776 and the Marie Skłodowska-Curie Individual 
Fellowship within the European Union’s Horizon 
2020 Program under Contract No. 706289.
J.B.M. acknowledges support of a Graduate Research Supplement for
Veterans from the Directorate for Mathematical and Physical Sciences
at the National Science Foundation.

\end{acknowledgments}


%

\end{document}